\documentclass[12pt,preprint]{aastex}
\pdfoutput=1
\shortauthors{Bower et al.}
\shorttitle{Black Hole Mass-Time Scale Correlation}
\begin{document}

\newcommand\degd{\ifmmode^{\circ}\!\!\!.\,\else$^{\circ}\!\!\!.\,$\fi}
\newcommand{\etal}{{\it et al.\ }}
\newcommand{\uv}{(u,v)}
\newcommand{\rdm}{{\rm\ rad\ m^{-2}}}
\newcommand{\msuny}{{\rm\ M_{\sun}\ y^{-1}}}
\newcommand{\mylesssim}{\stackrel{\scriptstyle <}{\scriptstyle \sim}}
\newcommand{\lsim}{\stackrel{\scriptstyle <}{\scriptstyle \sim}}
\newcommand{\gsim}{\stackrel{\scriptstyle >}{\scriptstyle \sim}}
\newcommand{\sci}{Science}
\newcommand{\sgr}{PSR J1745-2900}
\newcommand{\sgra}{Sgr~A*}
\newcommand{\kms}{\ensuremath{{\rm km\,s}^{-1}}}
\newcommand{\masy}{\ensuremath{{\rm mas\,yr}^{-1}}}

\def\kbar{{\mathchar'26\mkern-9mu k}}
\def\totd{{\mathrm{d}}}

%\slugcomment{Accepted for publication in the Astrophysical Journal}

\title{A Black Hole Mass-Variability Time Scale Correlation at
  Submillimeter Wavelengths}

\author{
Geoffrey C.\ Bower,\altaffilmark{1}
Jason Dexter,\altaffilmark{2}
Sera Markoff,\altaffilmark{3}
Mark A. Gurwell,\altaffilmark{4}
Ramprasad Rao,\altaffilmark{1}
Ian McHardy\altaffilmark{5}
}

\altaffiltext{1}{Academia Sinica Institute of Astronomy and Astrophysics, 645 N. A'ohoku Place, Hilo, HI 96720, USA; gbower@asiaa.sinica.edu.tw}
\altaffiltext{2}{Max Planck Institute for Extraterrestrial Physics, Giessenbachstr. 1, 85748 Garching, Germany}
\altaffiltext{3}{Anton Pannekoek Institute for Astronomy, University
  of Amsterdam, Science Park 904, 1098 XH Amsterdam, The Netherlands}
\altaffiltext{4}{Harvard-Smithsonian Center for Astrophysics, 60 Garden Street, Cambridge, MA 02138, USA}
\altaffiltext{5}{Department of Physics and Astronomy, The University, Southampton SO17 1BJ, UK}

\begin{abstract}
We analyze the light curves of 413 radio sources at submillimeter wavelengths
using data from the Submillimeter Array calibrator database.  The database
includes more than 20,000 observations at 1.3 and 0.8 mm that span 13
years.  We model the light curves as a damped random walk and determine
a characteristic time scale $\tau$ at which the variability amplitude
saturates.  For the vast majority of sources, primarily blazars and
BL Lac objects, we find only lower limits on $\tau$.  For two nearby 
low luminosity active galactic nuclei, M81 and M87, however, we 
measure $\tau=1.6^{+3.0}_{-0.9}$ days and $\tau=45^{+61}_{-24}$ days,
respectively ($2\sigma$ errors).  Including the previously measured
$\tau=0.33\pm 0.16$ days for Sgr A*, we show an approximately linear
correlation between $\tau$ and black hole mass for these nearby LLAGN.  
Other LLAGN with spectra that peak in the submm are expected to follow this
correlation. These characteristic time scales are comparable to 
the minimum time scale for emission processes close to an event
horizon, and suggest that the underlying physics may be independent of
black hole mass, accretion rate, and jet luminosity.

%This correlation demonstrates that submm emission in these sources
%arises from very close to the black hole event horizon and that the 
%underlying physics is independent of black hole mass, accretion rate, 
%and jet luminosity.
\end{abstract}


\section{Introduction}

In recent years there has been much work focused on understanding
black hole accretion and its similarities across the mass scale, from
stellar mass black holes in X-ray binaries (BHBs) to active galactic
nuclei (AGN).  Long term variability studies have found
evidence for a mass-dependence in timing features that holds from
BHB to AGN \citep{2006Natur.444..730M,2009ApJ...698..895K,2010ApJ...721.1014M}.  A linear
correlation can be understood
if the emission originates in a region that is the same size for all
systems in units of Schwarzschild radii, where
$R_S=2GM/c^2$.  However, the timescale is also inversely proportional
to the Eddington accretion fraction, suggesting that the pertinent
size of the emission zone is also regulated by the total system power.
Similarly, studies of broadband spectra have found a strong
correlation between radio and X-ray luminosity and black hole mass
(the ``Fundamental Plane of Black Hole Accretion'', or FP, see e.g.,
\citealt{2003MNRAS.345.1057M,2004A&A...414..895F,2012MNRAS.419..267P}),
that have confirmed earlier theoretical frameworks
\citep{1979ApJ...232...34B,1995A&A...293..665F} that synchrotron
spectral features scale predictably with black hole mass and accretion power.  
Combining these concepts, observations at a fixed
frequency of black holes of the same mass with different accretion
power should not yield similar timescales, because the frequency  ``selects''
out different sized emission regions in both cases.  

%submm band event horizon scales
The sub-millimeter (submm) band seems to be selecting
out regions of event-horizon scale in two sources of drastically
different mass and accretion rate: Sgr A* and M87
\citep{2008Natur.455...78D,2012Sci...338..355D,doelemanetal2012}. Both sources belong to the class of nearby, low-luminosity AGN (LLAGN;
\citealt{1999ApJ...516..672H}) that fall on the FP. 

%an independent test of size scale comes from variability: D+2014 for Sgr A*. Here we expand this analysis to a large sample of SMA calibrator light curves as an independent test of the size scale, and to compare LLAGN with regular AGN/blazars. 

Variability provides an independent test of the size scales. \citet{2014MNRAS.442.2797D} used submm
light curves to demonstrate that \sgra\ follows a damped-random
walk (DRW) variability pattern, with a characteristic time scale 
$\tau= 8_{-4}^{+3}$ h at 230 GHz at 95 per cent confidence,
with consistent results at higher frequencies. This time scale 
is an order of magnitude larger than the period of the last
stable orbit for a non-rotating black hole, which is most easily understood as resulting from accretion processes on a scale of a few to $\sim 10 R_S$. 
%and is most easily
%understood as the viscous time scale for the outer radius of
%the accretion disk on a scale of a few to $\sim 10 R_S$.

%, we present the first variability study of a
%sample of LLAGN in the submm, including these two sources and a
%``bridge'' source between them in mass and power, M81.  
%This builds on the work of 

We present a variability study of AGN in the submm, including the LLAGN M81 and M87 as well as the 411 other radio sources included in the SMA Calibrator database, which span more than a decade in duration (\S \ref{sec:data}). This sample was previously analyzed by \citet{2010arXiv1001.0806S}, but with a much smaller number of objects and total span. The full sample of light curves are well described by a noise process, which we model with a damped random walk in order to quantitatively measure characteristic variability time scales (\S \ref{sec:analysis}). We show that the LLAGN follow a correlation between black hole mass and time scale, while the  higher luminosity AGN have much longer characteristic time scales with no clear black hole mass dependence.

%In Section~\ref{sec:data}, we present the data.
%In Section~\ref{sec:analysis}, we present our time scale 
%analysis.  In Section~\ref{sec:conclusions} we show
%that the LLAGN follow a correlation between black hole mass
%and time scale, while the other, higher luminosity AGN have much
%longer characteristic time scales with no clear dependence on black hole mass.

\section{Data Sample \label{sec:data}}

The SMA calibrator database \citep{2007ASPC..375..234G} includes
observations of 412 sources from June 2002 to January 2015.
A total of 23254 flux densities are recorded with 19111 of
those flux densities obtained in the 1.3mm band.  
410 light curves have $>10$ flux densities, 141 light curves
have $>30$ flux densities, and 48 light curves have
$>100$ flux densities.  Observations were obtained at frequencies
from 200 to 406 GHz with 70\% obtained 
between 220 and 235 GHz. 165 observations (147 at 1.3mm) of 
M87 are included in the calibrator database spanning from January
2003 to January 2015.
Data reduction of sources followed standard SMA techniques
that set flux densities on absolute flux density scales determined
by solar system objects with typical accuracy of 5 to 10\%.

Sources are selected for the SMA calibrator database on the basis
of their suitability as phase reference sources for mm/submm 
interferometric observations, primarily reflected in 
their compact size, bright radio flux density, and declination
above -50$^\circ$.  Flux densities
in the data base range from 26 mJy to 52 Jy with a median value
of 1.2 Jy.  336 sources are matched to flat spectrum
radio sources in the CRATES catalog \citep{2007ApJS..171...61H}.
Non-matches to CRATES include steep spectrum calibrators
such as 3C 286 and sources in the Galactic plane such
as J1700-261 that may have been missed by the single dish surveys
that form the basis of CRATES.  The CRATES sample
includes primarily blazars and BL Lac sources;  289 of the sources
are found in the CGRABS catalog of $\gamma$-ray blazar
candidates \citep{2008ApJS..175...97H}.

We also included SMA monitoring observations of M81 in our analysis, which has been the target of a significant campaign (McHardy et al. 2015). This campaign
included 86 observations 
between September 2009 and March 2012, of which 42 observations
were obtained at frequencies higher than 300 GHz.  We also
include PdBI observations of M81 obtained in
2005 at 1.3 mm with a time resolution of $\sim 1$ hour \citep{2007A&A...463..551S}.  

\begin{figure}[p!]
\includegraphics[width=\textwidth]{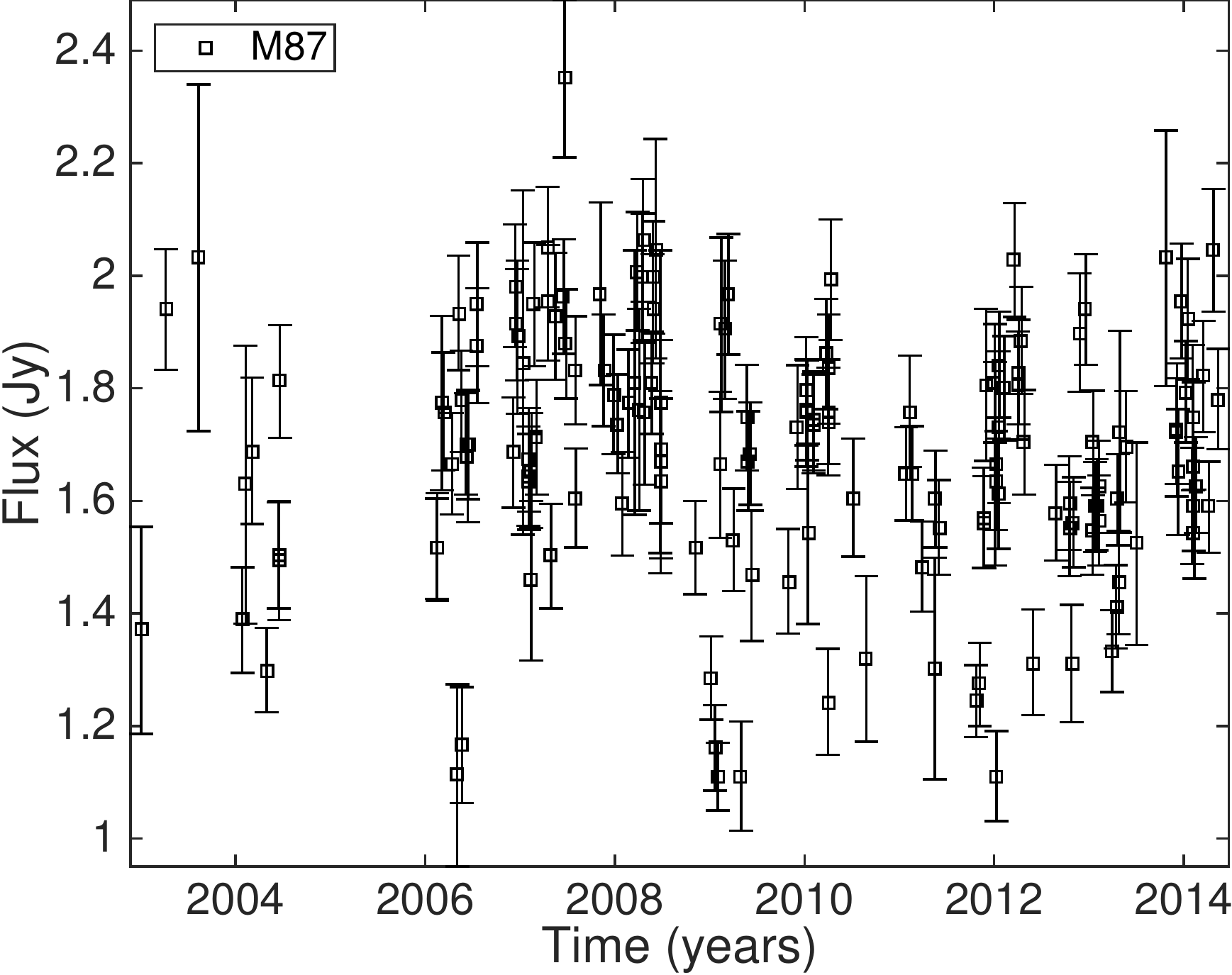}
\caption{
Light curve for M87 at 1.3 mm from the SMA.  
The light
curves show the characteristic of short-term variability coupled
with long-term stability that is the signature of the damped random
walk.
\label{fig:m87}
}
\end{figure}

\section{Time Scale Analysis \label{sec:analysis}}

\begin{figure}
\includegraphics[width=\textwidth]{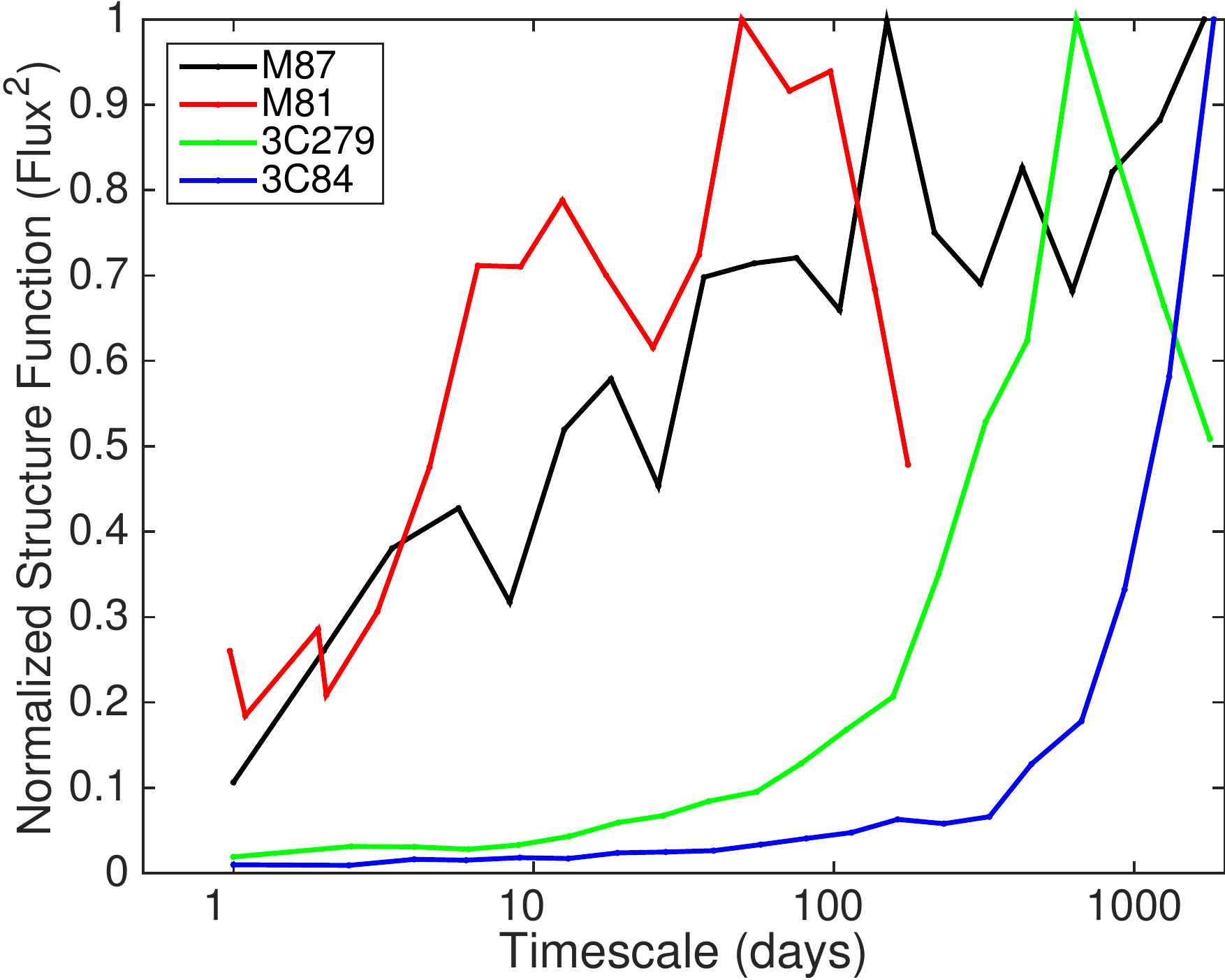}
\caption{
Structure functions for M87, M81, 3C 279, and 3C 84 from 1.3 mm SMA data.
The structure function suggests that M87 and M81 have different variability
characteristics from the two high power radio  sources.  Quantitative 
analysis with the DRW model confirms the differences.
\label{fig:sfun}
}
\end{figure}

The light curve for M87 is shown in Figure~\ref{fig:m87}.  The light
curve for M81 is shown in McHardy et al. (2015) and shares similar
characteristics of short-term variability and long-term stability. These properties are visible in the structure functions (Fig.\ref{fig:sfun}), defined for a light curve $s(t)$ with $N$ data points as 
\begin{equation}
S^2 (\Delta t) = {1 \over N} \Sigma\left( s(t) - s(t+\Delta t) \right)^2.
\end{equation}

Both structure functions show increases in activity on short time scales followed by a plateau, similar to those of \sgra\
\citep{2014MNRAS.442.2797D} but markedly different than those of
typical calibrators in the database.  For instance, the structure
functions for 3C 84 and 279 rise smoothly without any
apparent saturation time scale for $\Delta t < 10 \rm yr$.  
Structure functions are a useful
guide for qualitative comparison; however, they 
functions are unreliable for quantitative analysis
\citep{2010MNRAS.404..931E}. They can show artifacts from aliasing (e.g. peaks at $\sim 180$ days associated with the anual cycle), as well as spurious saturation time scales even with regularly sampled data. Fitting of structure functions for saturation time scales is also challenging since their values and errors at different time scales are correlated.

%The peaks in the M81 and M87 light curves occur at time scales 
%of $\sim 180$ days associated with the annual cycle.  Further, structure functions can
%show spurious saturation time scales even with regularly sampled data that span 10 times the time scale.

\subsection{Damped random walk model}

In order to quantitatively measure characteristic time scales, we
model the entire set of submm light curves as the result of a
stochastic damped random walk (DRW) process. 
The DRW is a simple 3-parameter model (mean, standard deviation of long-timescales,
and transition time) that is well-suited to parametrize the
most important properties of noise light curves
\citep{2009ApJ...698..895K}.  The DRW consists of red-noise on time scales
less than $\tau$ and white-noise on longer time scales.
The resultant structure function for the DRW is 
\begin{equation}
  S^2(\tau) = S^2_{\rm \infty} \left(1-e^{-|\Delta t|/\tau}\right), 
\end{equation}
\noindent where $S^2_{\infty}$ is the power in the light
curve on long time scales $\Delta t \gg \tau$. The characteristic time
scale $\tau$ determines the de-correlation time on which the
variability transitions from red noise on short timescales to white
noise on long time scales. This model has been shown to successfully
describe the optical light curves of quasars
\citep{2009ApJ...698..895K,2010ApJ...721.1014M}, as well as the submm light curve of
Sgr A* \citep{2014MNRAS.442.2797D}. 

We convert the likelihood of the observed light curves arising
from a given set of DRW parameters into the posterior probability of
the set of model parameters using a Bayesian approach. 
\citet{2009ApJ...698..895K} and \citet{2014MNRAS.442.2797D} used a
Metropolis-Hastings algorithm to sample the likelihood over the model
parameter space. We use this approach to measure the model parameters
in cases where the upper limit to $\tau$ is smaller than the light
curve duration $T$. However, this is not the case for the vast
majority of the SMA calibrator light curves. When no upper limit can
be found, the Metropolis-Hastings algorithm fails, as it
preferentially samples the long tail of the probability distribution
rather than its peak. 

To avoid this issue, we sample the probability distribution over a regular
$64^3$ grid in the parameters with uniform priors for the mean ($\mu$), standard deviation ($\log
\sigma$), and $\log \tau$. Since for the vast majority of sources we
are interested in lower limits on $\tau$, we choose the parameter
ranges to ensure that the lower end and peak of the
probability distribution are well captured, at the expense of the long
tails to large $\tau$ (much longer than the duration of the light
curve). From the probability
distributions in $\tau$, we consider good limits
to be cases where $\tau > \Delta t$ and $\tau < T$ at $99.7\%$
confidence, where $\Delta t$ is the minimum separation between two
measured points in the light curve and $T$ is the total light curve
duration. In all cases we report $2\sigma$ limits because of the non-Gaussian
nature of the distributions that can include long tails. 

Although less efficient than Monte Carlo, this direct grid method recovers consistent parameter estimates for the M81 and M87
light curves.  If the true light curve is not described by the DRW, as seen in Kepler quasar light curves \citep{2015MNRAS.451.4328K}, the resulting $\tau$ estimate will be biased. For light curves that can successfully be fit by the model ($\chi_\nu^2 \sim 1$), such as the SMA data used here, the bias leads to $< 1\sigma$ changes in our estimates of $\tau$.

%This is a conservative approach for measuring lower limits,
%since we are missing a small amount of the probability density in the
%tail. 

\subsection{Results}

\begin{figure*}[p!]
\begin{tabular}{cc}
\includegraphics[scale=0.5]{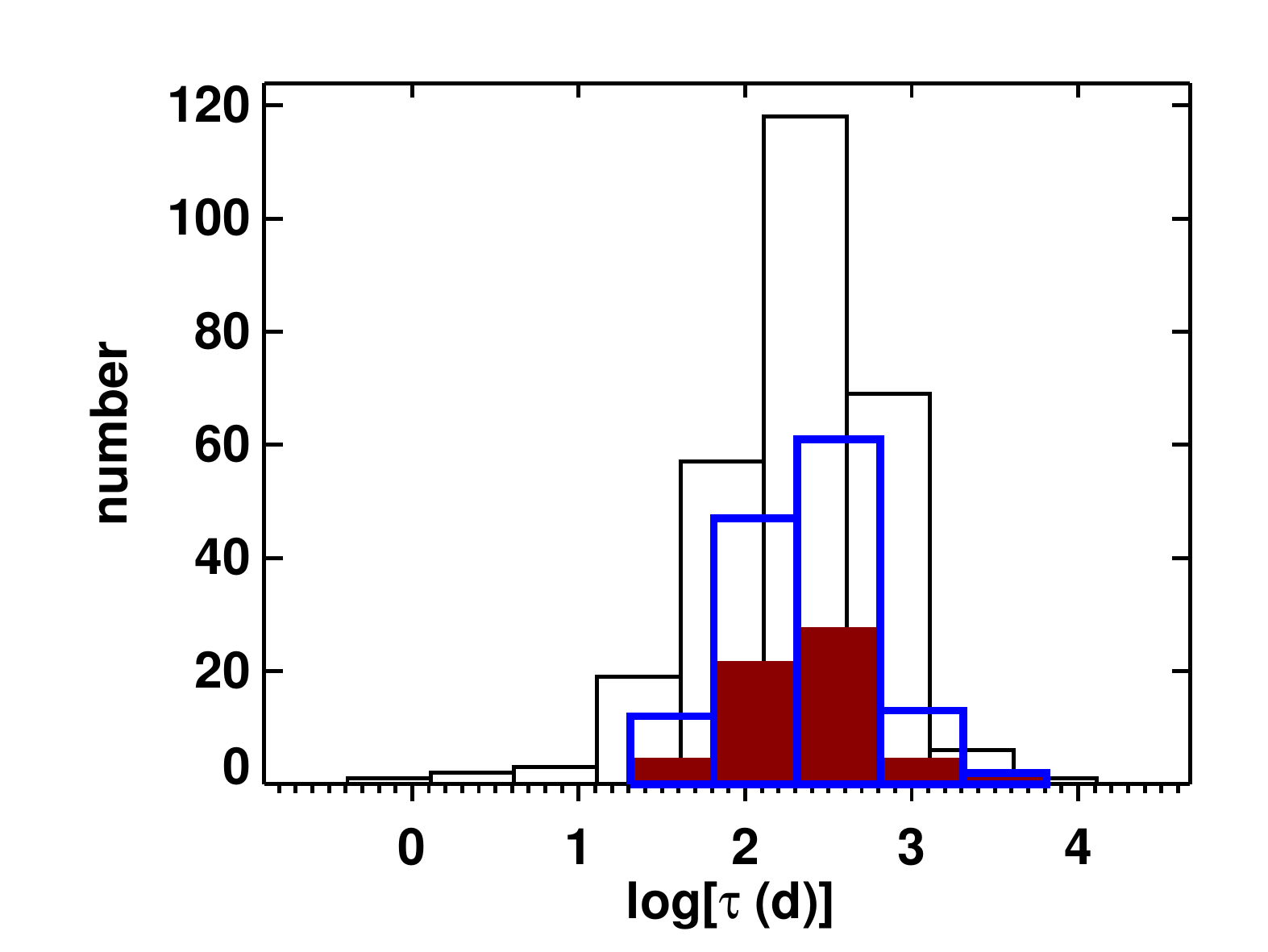}&
\includegraphics[scale=0.5]{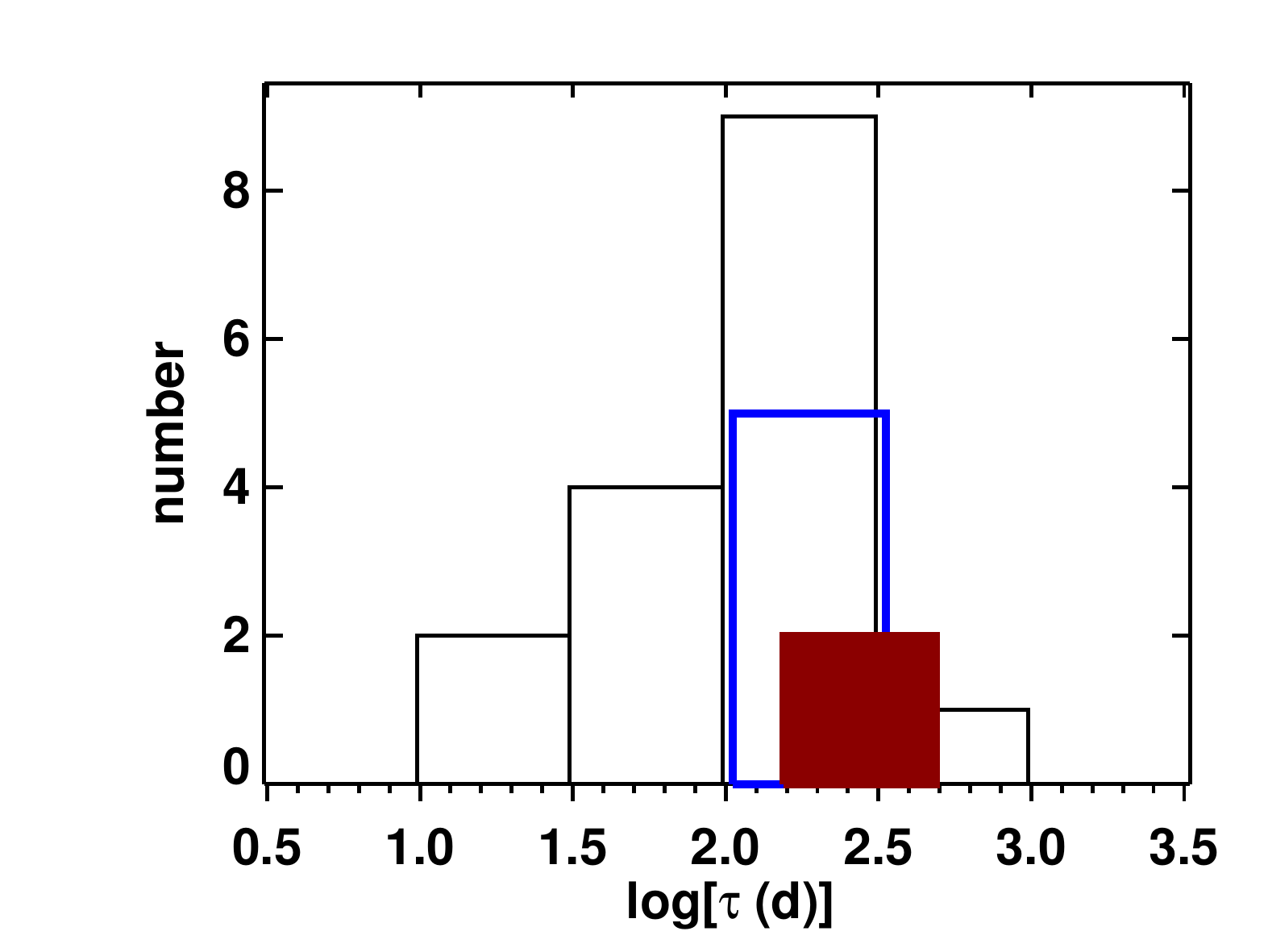}
\end{tabular}
\caption{
Histogram of $\tau$ lower (left) and upper (right) limits for entire
sample (black), those with $n > 30$ (blue),
and those with $n > 30$ and black hole mass measurements (red). Constraints
with $n < 30$ are used to avoid false positives. Those with black hole masses are representative of overall distribution. 
\label{fig:hists}
}
\end{figure*}

We use the DRW model to infer $\tau$ values from the
full set of 664 230 and 345 GHz light curves from the 413 sources. A
histogram of the resulting $95\%$ confidence lower and upper limits on $\tau$ is shown in Figure \ref{fig:hists}. In total, we find 276 light curves with lower
limits ($40\%$) and 16 with upper limits 
($2\%$). The histograms are further broken down by sources with $> 30$ data points and those with
available black hole mass estimates in the literature (see below). For
the measured lower limits, the sub-samples are consistent with the
full sample. Many of the measured upper limits are false positives in light curves with few data points caused by not sampling a large enough range in $\tau$.  
After those are removed, 5 detections of upper limits remain, all of which
have $\chi_\nu^2$ close to unity, similar to Sgr A*. In
Table~\ref{tab:detections}, we summarize the detections.
The light curve residuals after model subtraction are normally distributed for 
M81 and M87, as expected for an accurate model, and similar to what was seen for 
Sgr A* \citep{2014MNRAS.442.2797D}.

The results of this time scale analysis are in stark contrast for the
available LLAGN light curves (Sgr A*, M81, M87) and those of ordinary
AGN / blazars (the vast majority of the light curves). For the LLAGN,
3/3 sources have well measured values of $\tau$. For the rest of the
sources, $1\%$ (4/664) have reliable measurements of $\tau$. It is
therefore potentially possible to select LLAGN based on submm  properties alone,
with a low false positive rate and a high detection efficiency for
well sampled light curves. There is no systematic difference in the
number or cadence of measured flux densities that can explain this
result: it is related to an intrinsic difference in the submm
variability properties of the different source classes.

\begin{deluxetable}{lrrrrrl}
\tablecaption{Characteristic Time Scales for Submm Variability \label{tab:detections}}
\tablehead{
\colhead{Source} &
\colhead{$\tau$} &
\colhead{$\tau_{\rm low}$} &
\colhead{$\tau_{\rm high}$} &
\colhead{$\chi_\nu^2$} &
\colhead{$M_{BH}$} &
\colhead{Source}
\\
& 
 \colhead{(days)} &
 \colhead{(days)} &
 \colhead{(days)} &
 &
\colhead{($M_\sun$)}
}
\startdata
0433+053 &    106  &    63 &      299 & 0.88 & $2.6 \times 10^7$ & \citet{2002ApJ...579..530W} \\
0721+713 &    75   &   56 &      155 & 0.83 & $5.2 \times 10^6$ & Fundamental Plane \\
1104+382 &    40   &   16 &      154 & 0.81 & $1.9 \times 10^8$ &\citet{2002ApJ...579..530W} \\
1751+096 &    106 &     50 &      331 & 1.04 & $2.2 \times 10^7$ & Fundamental Plane\\
M87      &     45  &    21 &      106 & 0.87 & $6.2 \times 10^9$ & \citet{2013ARAA..51..511K} \\
M81      &      1.6 &  0.7  &  4.6 & 1.14 & $6.5 \times 10^7$ &    \citet{2009MNRAS.394..660C} \\
\hline
Sgr A*\tablenotemark{1}  &  0.33 &  0.17 &  0.5 & 1.05 & $4.4 \times 10^6$ & \citet{2010RvMP...82.3121G} \\
\enddata
\tablenotetext{1}{\citet{2014MNRAS.442.2797D}}
\end{deluxetable}

\subsection{Black Hole Mass Estimates}

We compiled black hole masses from the literature.  The three nearby LLAGN have the most accurately determined
masses, often from multiple dynamical methods.  \sgra\ has a mass
of $4.4 \times 10^6 M_{\sun}$ \citep{2010RvMP...82.3121G}.
M87 has a mass of $3.5^{+0.9}_{-0.7} \times 10^9\, M_\sun$ from gas dynamic
measurements \citep{2013ApJ...770...86W} and a mass of $6.6 \pm  0.4 \times 
10^9\, M_\sun$ from stellar dynamic measurements \citep{2011ApJ...729..119G}.
Following \citet{2013ARAA..51..511K}, we adopt a value of $6.2 \times 10^9\,
M_\sun$.  Similarly for M81, stellar and gas dynamical measurements
exist \citep{2000AAS...197.9203B,2003AJ....125.1226D} and we follow \citet{2013ARAA..51..511K} to adopt a value of $6.5 \times 10^7\, M_\sun$.

%Centaurus A, which is a transitional object between LLAGN and high luminosity systems \citep{2004ApJ...612..786E},
%has a well-determined black hole mass of $6 \times 10^7\, M_\sun$ 
%\citep{2009MNRAS.394..660C,2013ARAA..51..511K}.

For the remainder of the sample, we rely on published black hole mass estimates from various techniques.  These
include broad line region spectroscopy \citep{2002ApJ...579..530W,2007ApJS..168....1K,2008AJ....135..928S,2011ApJS..194...45S,2014A&A...563A..54P},
black hole mass-bulge luminosity correlation
\citep{2002ApJ...579..530W}, and the FP between
radio-X-ray luminosity
\citep{2003MNRAS.345.1057M,2012MNRAS.419..267P}.  For the case of
\citet{2014A&A...563A..54P}, which does not provide mass estimates, we
use their measurements of the Mg II transition and the methodology of
\citet{2011ApJS..194...45S} to estimate masses. 
FP estimates of the black hole mass are
derived using ROSAT 0.2 - 2.0 keV X-ray fluxes and the 1.3mm flux
density that is debeamed assuming a Doppler boosting factor of 7.  We
identify 194 black hole mass measurements for 148 sources.  For
sources with multiple mass estimates, we rely on the most recent
published estimate.  The accuracy of the mass estimates range from 0.16 dex 
for Mg II transitions to $>0.4$ dex for continuum luminosity \citep{2011ApJS..194...45S}.

\begin{figure}[p!]
\includegraphics{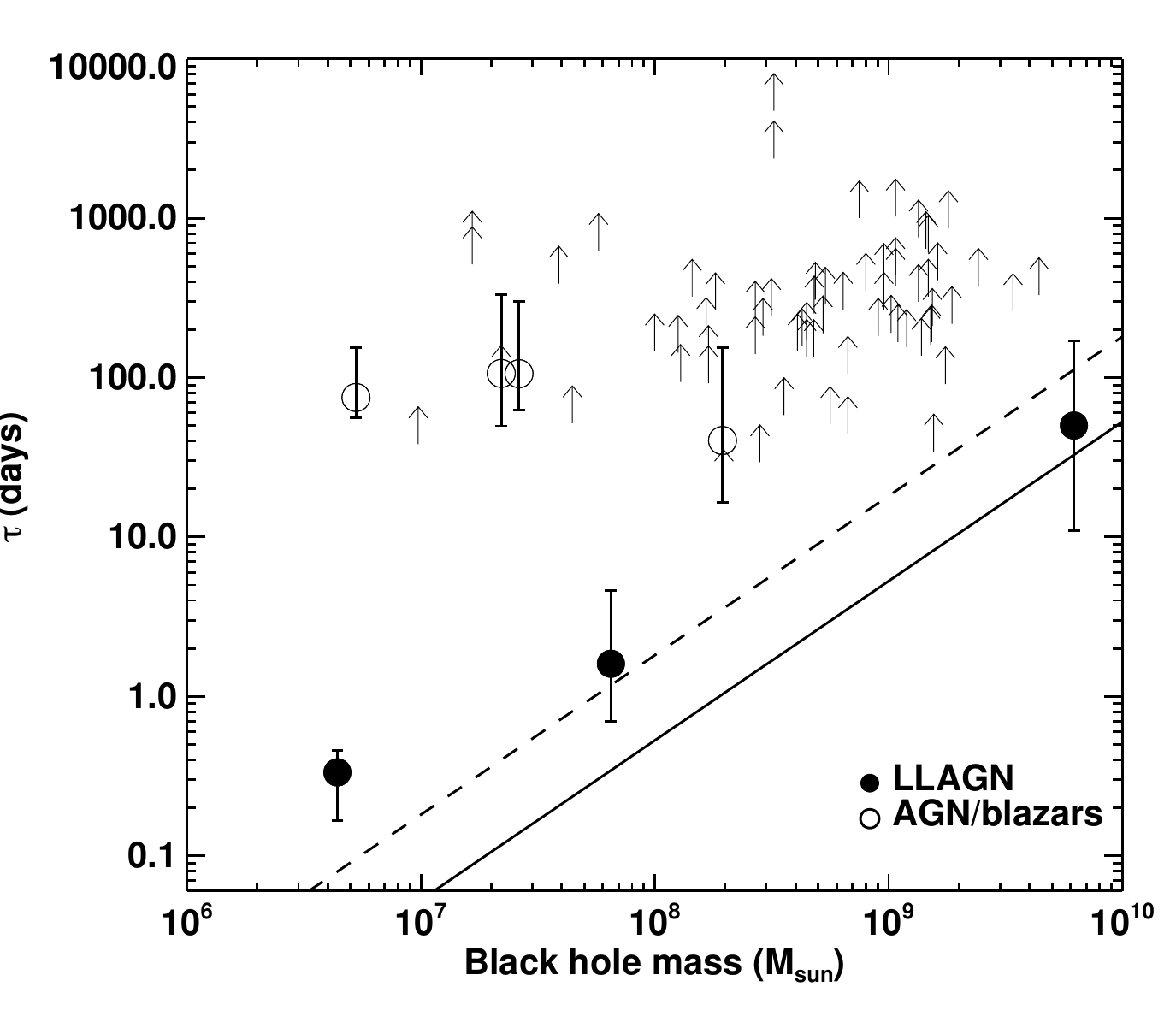}
\caption{
Black hole mass versus DRW time scale for the SMA calibrator source sample.  LLAGN Sgr A*, M81,
and M87 are shown as filled in circles.  Other
AGNs from the SMA calibrator database with black hole masses are shown
as open circles for detections and up arrows for lower limits on $\tau$. 
The time scales associated with AGN/blazars are
significantly longer at the same black hole mass than those of the
LLAGN. The inferred time scales for LLAGN variability increase with
black hole mass, and are consistent with the predicted linear
relationship from General Relativity.  The solid line gives the period
for the innermost stable circular orbit (ISCO); the dashed line gives the
infall time for a disk with radius $5 R_S$, a viscosity $\alpha=1$, and a
height $H/R=1$.
\label{fig:taumbh}
}
\end{figure}

\section{Discussion \label{sec:conclusions}}

The $\tau$-$M_{\rm BH}$ relationship for the sub-sample of SMA
calibrator sources with black hole mass estimates is shown in Figure
\ref{fig:taumbh}. The difference in submm variability properties
between regular and low-luminosity AGN is immediately apparent: the
vast majority of AGN only yield lower limits on $\tau$, while the
LLAGN have measured time scales that lie well below many of these
 at comparable black hole mass.

On the other hand, the measured $\tau$ values for the 3 LLAGN with
good measurements of $\tau$ do show a significant trend of 
increasing $\tau$ with increasing black hole mass.  Formally, the
best power-law fit is
$\tau \approx 0.3 {\rm\, day\,} (M_{\rm BH}/4 \times 10^6 M_\odot)^{0.7 \pm 0.1}$
(with a coefficient of determination $R^2=0.996$).
We consider the observed trend consistent with a linear relationship, however,
given the small number of sources and the large uncertainties in 
$\tau$ and $M_{\rm BH}$.  If the M87 mass were taken from gas dynamical measurements,
 for instance, then a linear law would be consistent with the data. 
The reliability of the correlation rests on only three LLAGN.  This is a minimally small number of objects for a correlation, 
but these are the only LLAGN for which $\tau$ can currently be determined.  
We are performing observational campaigns to increase the number of 
LLAGN with measured $\tau$.

A prediction of general relativity is that time scales around black
holes should scale as $\tau \sim R^{3/2} M_{\rm BH}$, where $R$ is the radius in
units of the event horizon size. This time scale reaches a minimum for
a given $M_{\rm BH}$ when $R$ is comparable to the size of the event
horizon, and in that case one might expect to find a
linear relationship between $\tau$ and $M_{\rm BH}$. This is clearly
not borne out by the full sample: there is a large scatter in measured
$\tau$ values for regular AGN, and they do not fall on a linear track
with the LLAGN. The discovery of such a relationship for the LLAGN
suggests that we are measuring such minimum time scales from emission
close to an event horizon.

This interpretation is consistent with our understanding of these sources inferred from 
their spectral energy distributions.  For all three LLAGN, the spectra peak in the millimeter
or submillimeter regime \citep{2015ApJ...802...69B,2008ApJ...681..905M,2013EPJWC..6108008D}. Synchrotron self-absorption leads to a large photosphere which shrinks
until the spectral peak,  where we can see down to the event
horizon. At longer wavelengths, we expect longer timescales because of the
larger scales of the photosphere. This larger photosphere is seen as a growing
intrinsic image size with wavelength in VLBI
observations of Sgr A* \citep{2006ApJ...648L.127B} and M87
\citep[e.g.,][]{2011Natur.477..185H}. The same effect shows up in the
variability time scales for Sgr A* and M81
and likely for M87.  For Sgr A*,
\citet{2006ApJ...641..302M} found a characteristic time scale at wavelengths
from 0.7 to 20 cm that ranged from 6 days to hundreds of days, increasing
with wavelength.  Additionally, we performed a  DRW analysis of the 2cm M81 light
curve \citep{1999AJ....118..843H} and 
found a lower limit to $\tau$ of tens of days, much larger than the
measured value at 1.3mm.
In addition, for the case of Sgr A*, the NIR variability time scale 
matches the submm value, consistent with the hypothesis of emission
at both wavelengths emerging from the same region near the 
event horizon \citep{2012ApJS..203...18W}.  Thus, one should expect a similar
relationship for other LLAGN with spectra peaking in the submm.
The blazars and high power radio sources presumably do not follow a linear correlation
because these sources are still optically thick at this wavelength.
X-ray binaries in the low/hard state may also show the same correlation in the optically
thin regime, although this
may be difficult to observe due to the $\lsim 1$ sec time scales implied by the relationship
and the rapidly evolving dynamics of these sytems \citep{2014MNRAS.439.1390R}.

%Cen A is also part of the SMA calibrator database. Its luminosity is substantially
%sub-Eddington but it is considerably more luminous than even M87 \citep{2004ApJ...612..786E},
%hence, it is not considered an LLAGN.  Nevertheless, as a nearby, unbeamed source with a well-determined
%black hole mass, it is worth a closer look.  We
%find a tentative $2\sigma$ lower limit of $\sim 27$ days.  However,
%only 23 flux densities were obtained over $>3000$ days
%and the DRW fit is poorly constrained. If this lower limit for $\tau$ is accurate,
%then Cen A, which has a mass comparable to that of M81, does not fall
%on our correlation.  This may be the consequence of a 
%spectrum that peaks between $10^{13}$ and $10^{14}$ Hz
%\citep{2010A&A...511A..64V}, where the submm emission likely comes from a
%larger radius.

A similar relationship was found in the X-ray 
\citep{2006Natur.444..730M} for black holes ranging from stellar to
supermassive. This relationship, however, relied on a scaling factor
based on the accretion rate, which we do not require here. This scaling factor could be interpreted as compensating between different
values of $R$ in the time scale corresponding to the X-ray emission region. In the
LLAGN, we instead appear to have reached the minimum variability time scale
corresponding to emission from event horizon scales around black
holes. 
%It is possible that including a similar scaling factor could
%allow the AGN/blazars to fall on the same correlation as the LLAGN. 

The scaling relationship between black hole mass and variability time scale for
LLAGN is an important insight for Event Horizon Telescope observations
of these sources \citep{2013arXiv1309.3519F}.  
These imaging observations will have resolutions as good as a few Schwarzschild radii
and have the potential to probe fundamental 
gravitational physics \citep[e.g.,][]{2014ApJ...788..120L,2014ApJ...784....7B,2014MNRAS.445.3370P,2015MNRAS.446.1973R}.  
The $\tau-M_{BH}$ correlation shows that the compact sizes measured for 
these sources are actually colocated with the black hole.
In addition, many of the fundamental parameters such as spin are degenerate  with
source physics parameters including the accretion rate, the jet inclination
angle, the electron temperature distribution, the
magnetic field strength and orientation, and the optical depth.
Given the vastly different scales, environments, and physical properties 
of these three LLAGN systems, the existence of this correlation demonstrates
a surprising coherence among these sources. These,
and potentially other LLAGN with spectra peaking in the mm/submm can be treated
as a unified class.  Broadband spectra and light curve monitoring of other nearby LLAGN 
in the submm will add to the number of sources that fall into this class.
Sources that do not follow this correlation, such as the majority of the
SMA calibrators, may follow the relation at a higher frequency where the
emission becomes optically thin.

\acknowledgements
The Submillimeter Array is a joint project between the Smithsonian Astrophysical Observatory and the Academia Sinica Institute of Astronomy and Astrophysics and is funded by the Smithsonian Institution and the Academia Sinica.
This research has made use of the NASA/IPAC Extragalactic Database (NED), which is operated by the Jet Propulsion Laboratory, California Institute of Technology, under contract with the National Aeronautics and Space Administration.

%\bibliographystyle{apj}
%\bibliography{myrefs,refs}

\end{document}